\pdfoutput=1
\documentclass[aps,10pt,showpacs,floatfix,nofootinbib,twocolumn,superscriptaddress]{revtex4-1}
\usepackage{subfigure}
\usepackage{amssymb}
\usepackage{amsfonts}
\usepackage{amsmath}
\usepackage{amsthm}
\usepackage{epsfig}
\usepackage{graphicx}
\usepackage[usenames,dvipsnames]{color}
\usepackage[latin1]{inputenc}
\usepackage{hyperref}
\usepackage{comment}

\begin{document}
\title{Unexpected validity of Schottky's conjecture for two-stage field emitters: a response via Schwarz-Christoffel transformation}

\date{\today}

\author{Edgar Marcelino}
\email{edgarufba@gmail.com}
\address{Centro Brasileiro de Pesquisas F\'{i}sicas, Rua Dr. Xavier Sigaud 150, 22290-180, Rio de Janeiro, RJ, Brazil}
\address{Instituto de F\'{\i}sica, Universidade Federal da Bahia,
   Campus Universit\'{a}rio da Federa\c c\~ao,
   Rua Bar\~{a}o de Jeremoabo s/n,
40170-115, Salvador, BA, Brazil}

\author{Thiago A. de Assis}
\email{thiagoaa@ufba.br}
\address{Instituto de F\'{\i}sica, Universidade Federal da Bahia,
   Campus Universit\'{a}rio da Federa\c c\~ao,
   Rua Bar\~{a}o de Jeremoabo s/n,
40170-115, Salvador, BA, Brazil}

\author{Caio M. C. de Castilho}
\email{caio@ufba.br}
\address{Instituto de F\'{\i}sica, Universidade Federal da Bahia,
   Campus Universit\'{a}rio da Federa\c c\~ao,
   Rua Bar\~{a}o de Jeremoabo s/n,
40170-115, Salvador, BA, Brazil}
\address{Instituto Nacional de Ci\^{e}ncia e Tecnologia em Energia e Ambiente - INCTE\&A, Universidade Federal da Bahia,
   Campus Universit\'{a}rio da Federa\c c\~ao,
   Rua Bar\~{a}o de Jeremoabo s/n,
40170-280, Salvador, BA, Brazil}

\begin{abstract}

The electric field in the vicinity of the top of an emitter with a profile consisting of a triangular protrusion on an infinite line is analytically obtained when this system is under an external uniform electric field. The same problem is also studied when the profile features a two-stage system, consisting of a triangular protrusion centered on the top of a rectangular one on a line. These problems are approached by using a Schwarz-Christoffel conformal mapping and the validity of Schottky's conjecture (SC) is discussed. We provide an analytical proof of SC when the dimensions of the
upper-stage structure are much smaller than the lower-stage ones, for large enough aspect ratios
and considering that the field enhancement factor (FEF) of the rectangular structure is evaluated
on the center of the top of the structure, while the FEF of the triangular stage is evaluated near
the upper corner of the protrusion. The numerical solution
of our exact equations shows that SC may remain valid even when both stages feature dimensions
from the same order of magnitude, reinforcing the validity of SC for multi-stage field emitters.
\end{abstract}

\maketitle

\section{Introduction}

The potential barrier experienced by the electrons in Cold Field Emission (CFE) \cite{Jeffreys,FowlerN,Burgess,MG,Forbes,ForbesJPhysA,Forbes2013,Cole2015chapter} is dependent of the local electrostatic field in the surface of the emitter, which can be estimated from analytical solutions of Laplace equation for specific boundary conditions, depending on the shape of the emitter and the applied field. Thus, obtaining these classical solutions and understanding their general features is an important task in order to describe CFE, and consequently for proposing useful applications to electronic devices. The applied electric field, required to produce field emission in pure metals with a smooth surface, is typically of the order of a few V/nm. For practical purposes, it is useful to achieve high local electric fields by considering geometries that feature a great local enhancement of the applied electric field (typically between $10^2$ and $10^3$). This can be done by considering shapes containing corners, edges and tips with a high aspect ratio (ratio height/width) \cite{Edgcombe}. A simple example consists on a metal single tip field emitter (STFE). Alternatively one could consider a two-stage-structure which, depending on the dimensions of each of the stages, produces a high local field enhancement factor (FEF), as compared with the single-stage case \cite{LiuNew,Huang,Jensen,deAssis1}. All these aspects have motivated the development of field emitter designs featuring these properties \cite{Shiffler1,Shiffler2,Jones,deAssisJAP}.

Following the aforementioned motivations, some recent works \cite{Ryan1,Ryan2,Jensen,deAssis1,Jensen2017} have studied the validity of Schottky's conjecture (SC) \cite{Schottky}. This conjecture states that the FEF of a two-stage field emitter is the product of the FEFs of each of the two stages. SC became a paradigm of a good approximation in order to explain large FEFs obtained in CFE experiments, nevertheless it is well known that this conjecture is not rigorously true and there is still some lack of theoretical results studying its validity. Previous results suggest that SC is valid when the dimensions of the upper-stage structure are much smaller than the ones from the lower-stage \cite{Ryan1,Ryan2,Jensen,deAssis1}. Furthermore, results using a point charge model suggest that SC remains valid even for some situations in which the two stages have dimensions of the same order of magnitude \cite{Jensen}. In this work, we explore the validity of SC for specific geometries.

In order to investigate these aspects, we have considered a two-dimensional emitter consisting of a conducting line with a conducting triangular protrusion under an applied uniform electrostatic field and evaluate the FEF at the vicinity of the apex (top corner) of the emitter. The same problem is also extended to the case in which the line has a triangular protrusion centered on the top of a rectangular one and the validity of SC \cite{Schottky} is investigated through the whole space of parameters describing the geometry of the emitter. Thus, the present work focuses on ridge emitters. Although CFE experiments involve three-dimensional geometries, ridge emitters can be used to provide both numerical and analytical solutions that can be used to understand SC in a broader sense, which is certainly relevant to achieve a better comprehension on this topic. Our problems are approached by using the Schwarz-Christoffel conformal mapping \cite{Churchill, Hilderbrand}.

This paper is organized as follows. In Sec. II, the FEF
in the vicinity of the top of a conducting triangular protrusion on a flat line is analytically derived.
In Sec. III, this problem is extended to a two-stage structure formed by a triangular protrusion placed on the top of
a rectangular one on a conducting line. An analytical proof of SC is presented when the dimensions of the lower protrusion are much larger than the ones from the other and for high aspect ratios.
Numerical results are presented exploring the whole space of parameters that characterize the emitter's geometry and the validity of SC under other limits is discussed. In Sec. IV, we summarize our results and present our conclusions.

\section{Single-stage structure}

\begin{figure*}[ht]
	\centering    
	\hspace{-1.0cm}
    \includegraphics[width=0.26\linewidth]{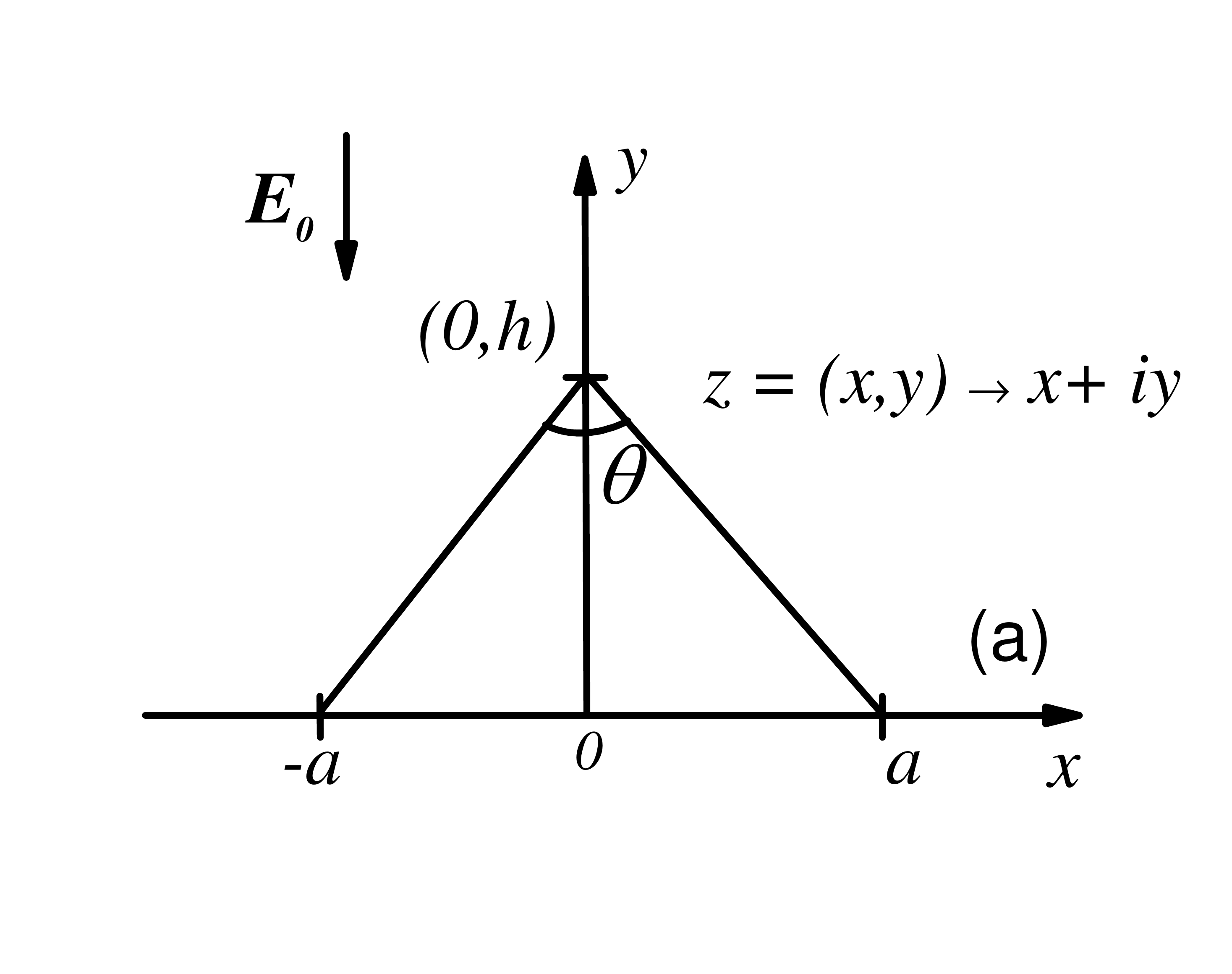}
   \hspace{-1.0cm}
    \includegraphics[width=0.26\linewidth]{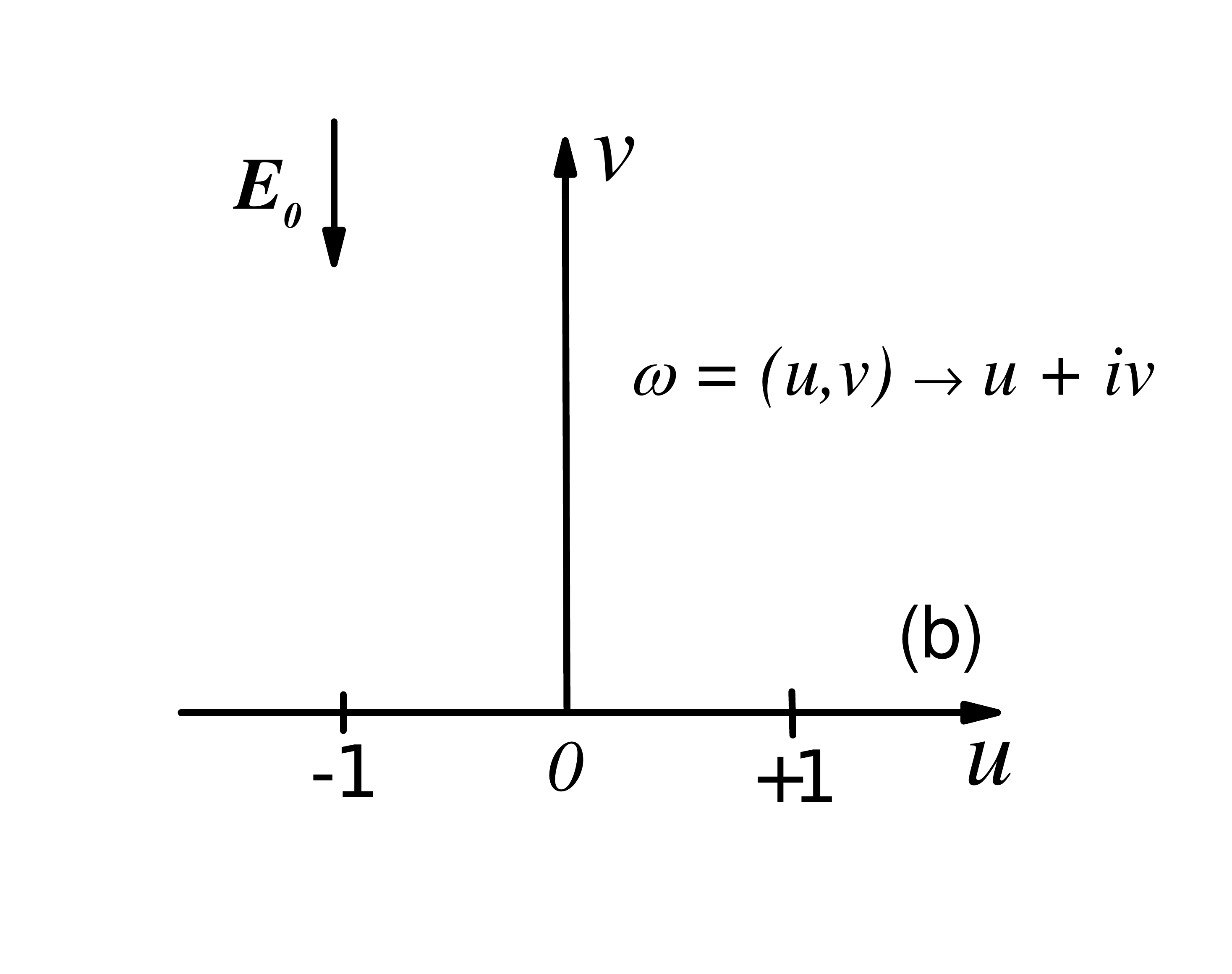}
    \includegraphics[width=0.27\textwidth]{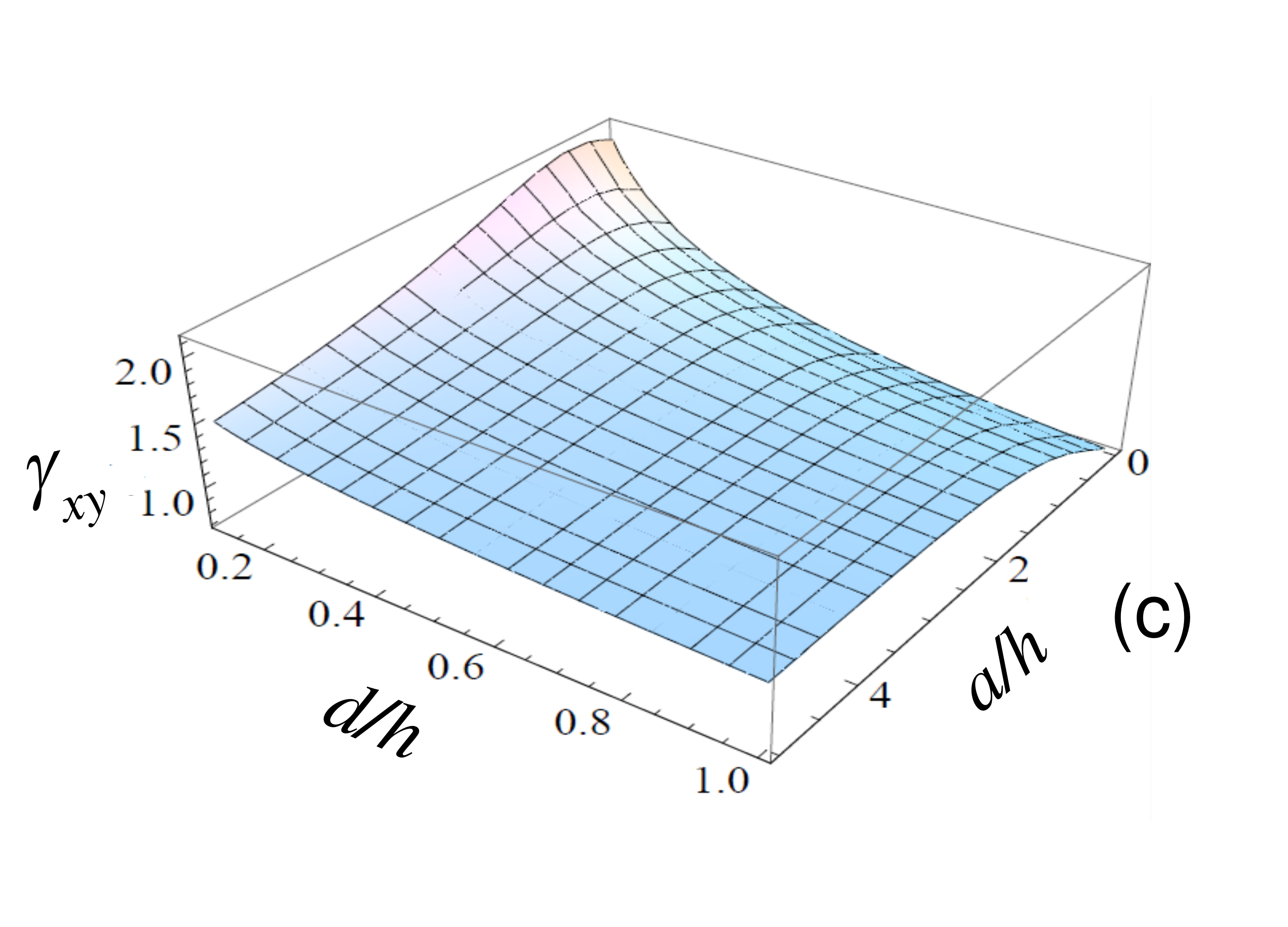}
      \includegraphics[width=0.27\textwidth]{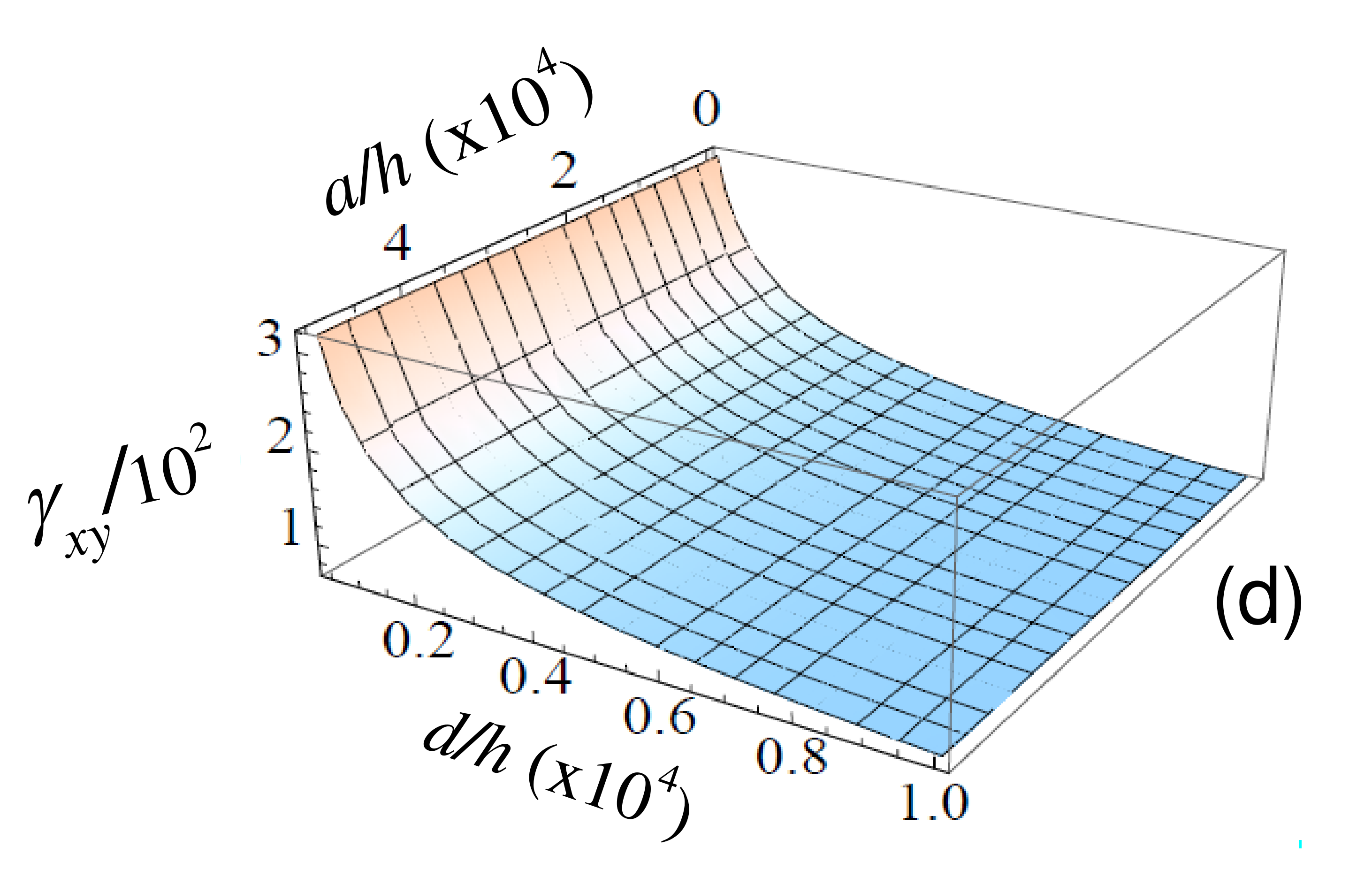}
	\caption{Conformal mapping between the (a) $z$ and (b) $w$ planes for the two-dimensional problem of a field emitter consisting of a triangular protrusion of height, $h$, and half-width, $a$, on a conducting line. The system is under an external electrostatic field $\mathbf{E_{0}}$. (c) FEF, $\gamma_{xy}$, calculated from Eq.(\ref{Enhancement1}), as a function of the ratios $a/h$ and $d/h$. (d) Inset from (c), showing larger values of FEF, useful for technological applications.}
	\label{Trianglew}
\end{figure*}

Let us consider first the two-dimensional problem of a field emitter consisting of a triangular protrusion of height $h$ and half-width $a$ on a flat surface (line) under an external electric field $\mathbf{E_{0}}$, as it is showed in Fig.\ref{Trianglew}(a). We obtain the electric field under electrostatic equilibrium for this system by conformal mapping between $w$ and $z$ planes, as showed, respectively in Figs.\ref{Trianglew}(b) and \ref{Trianglew}(a), where $w=(u,v)=u+iv$ and $z=(x,y)=x+iy$. This mapping transforms the $u$-axis in the $w$-plane into the polygonal line showed in Fig.\ref{Trianglew}(a) and it is such that the following correspondences are fulfilled: $w=(\pm 1,0) \leftrightarrow z=(\pm a,0)$ and $w=(0,0) \leftrightarrow z=(0,h)$. This is a particular case of a Schwarz-Christoffel transformation, which can be written in the following way:
\begin{equation}   \label{SC}
z(w)=A \int_{z_{0}}^{w} f(w) dw + B,
\end{equation}
where $z_{0}$ is arbitrary and the function $f(w)$ is determined from the vertices and angles of the polygonal line in which the $u$-axis is mapped, i.e.,
\begin{equation}
f(w)=\frac{w^{1-\alpha}}{(w^{2}-1)^{\frac{1-\alpha}{2}}}.
\end{equation}
The angle $\theta$ of the triangular protrusion is given by $\theta=\alpha \pi$ $(0<\alpha<1)$, as shown in Fig.\ref{Trianglew}(a). This definition of the $\alpha$-parameter is useful for writing the conformal mapping in a simple way.

Considering $z_{0}=0$ and that $z(w=0)=ih$, it is straightforward to obtain $B=ih$. The last constant in the Schwarz-Christoffel transformation, $A$, can be determined from the correspondence $z(w=\pm1)=\pm a$, which leads to the following equation
\begin{equation} \label{condition_A}
a-ih=A \int_{0}^{1}f(w)dw=\frac{A}{2} e^{i \left( \frac{\theta}{2}-\frac{\pi}{2} \right)} B \left(1-\frac{\alpha}{2},\frac{1+\alpha}{2} \right).
\end{equation}

The integral representation of the ``Beta" function used in Eq.(\ref{condition_A}), $B(a,b)$, can be written as
\begin{equation}
B(a,b)=\frac{\Gamma(a) \Gamma(b)}{\Gamma(a+b)}=\int_{0}^{1} x^{a-1}(1-x)^{b-1}dx,
\end{equation}
where $\Gamma(s) \equiv \int\limits_{0}^{\infty} e^{-x} x^{s-1} dx$ is the gamma function. From Eq.(\ref{condition_A}) one can easily obtain $\tan \left(\frac{\theta}{2} \right)=\frac{a}{h}$, which is obviously true from Fig.\ref{Trianglew}(a), and also determine $A$ as follows:
\begin{equation}
A=\frac{\sqrt{\pi(a^{2}+h^{2})}}{ \Gamma \left(1-\frac{\alpha}{2}\right) \Gamma \left(\frac{1+\alpha}{2} \right)}.
\end{equation}
Thus, the mapping between $w$ and $z$-planes is given by
\begin{equation} \label{Mapping1}
z=\frac{\sqrt{\pi} (a-ih)}{ \Gamma \left( 1-\frac{\alpha}{2} \right) \Gamma \left( \frac{1+\alpha}{2} \right)}  \int_{0}^{w} \frac{w^{1-\alpha}}{(1-w^{2})^{\frac{1-\alpha}{2}}}dw +ih.
\end{equation}
The complex electrostatic potential, $\phi(w)=i A |\mathbf{E_0}| w$, yielding an uniform electric field in the $w$-plane, can be used to determine the complex electrostatic field solution in the $z$-plane,
\begin{equation} \label{Field1}
E_{x}-iE_{y}=\frac{d \phi}{dz}=\frac{d \phi /dw}{dz/dw}=\frac{i E_{0}}{f(w)}.
\end{equation}
Near the corner $w=0 \Rightarrow z=ih$ in the $w$-plane, we may consider $f(w) \approx e^{-i \left(\frac{\pi}{2}-\frac{\theta}{2} \right)} w^{1-\alpha}$. Thus, Eq.(\ref{Mapping1}) becomes,
\begin{equation}
|z-ih| \approx \frac{\sqrt{\pi} |a-ih| |w|^{2-\alpha}}{(2-\alpha) \Gamma \left( 1-\frac{\alpha}{2} \right) \Gamma \left( \frac{1+\alpha}{2}\right)}.
\label{Fieldn}
\end{equation}
Equations (\ref{Field1}) and (\ref{Fieldn}) can be used to determine the FEF at the vicinity of the corner $(w=0)$ of the triangular protrusion. This local FEF, $\gamma_{xy}$, is given by
\begin{equation} \label{Enhancement1}
\gamma_{xy} \equiv \frac{|\mathbf{E}(x,y)|}{E_0} \approx \left[\frac{\sqrt{\pi (a^{2}+h^{2})}}{(2-\alpha) \Gamma \left(1-\frac{\alpha}{2}\right) \Gamma \left( \frac{1+\alpha}{2} \right) d} \right]^{\frac{1-\alpha}{2-\alpha}},
\end{equation}
where $d \equiv d(x,y) \equiv \sqrt{x^{2}+(y-h)^{2}}$ is the distance to the corner ($w = 0$) where we have evaluated the FEF.

In Fig.\ref{Trianglew}(c), the FEF, $\gamma_{xy}$, is plotted as a function of the ratios $a/h$ and $d/h$. It is interesting to notice that, for a fixed distance from the corner, there is an optimized ratio $a/h$, corresponding to an optimized angle $\theta$, yielding the largest FEF. This optimized angle tends to zero as long as $d  \rightarrow 0$, yielding an infinity FEF on the corner $(d=0)$ [see Fig.\ref{Trianglew}(d)]. An important angle dependence in the FEF has also been found in a previous work, considering a triangular protrusion inside an infinite channel \cite{Venkattraman}. However, we stress that the present work is focused in CFE. Thus, the gap-length between the emitter and the counter electrode is much greater than the height $h$. This limit is expected to be obeyed, for analytical models, by considering an infinite gap-length. In this case, the FEF, $\gamma_{xy}$, is not expected to be dependent on the applied field $\mathbf{E_{0}}$ \cite{Qin,Forbes2012b}, as showed in Eq.(\ref{Enhancement1}).

\section{Two-stage structure}

\begin{figure*}[ht]
	\centering    
    \hspace{-1.0cm}
    \includegraphics[width=0.30\linewidth]{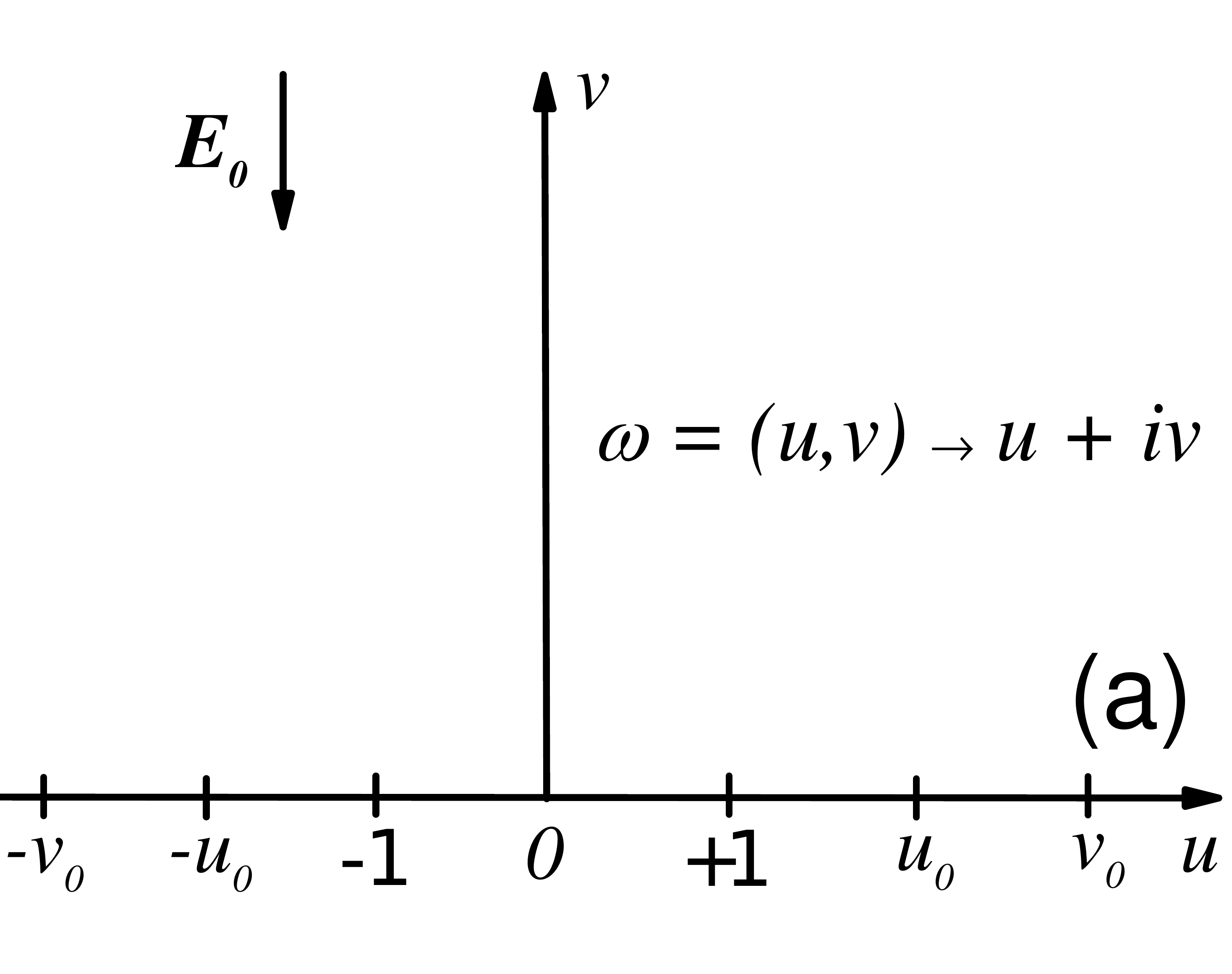}
    \hspace{+0.3cm}
    \includegraphics[width=0.30\linewidth]{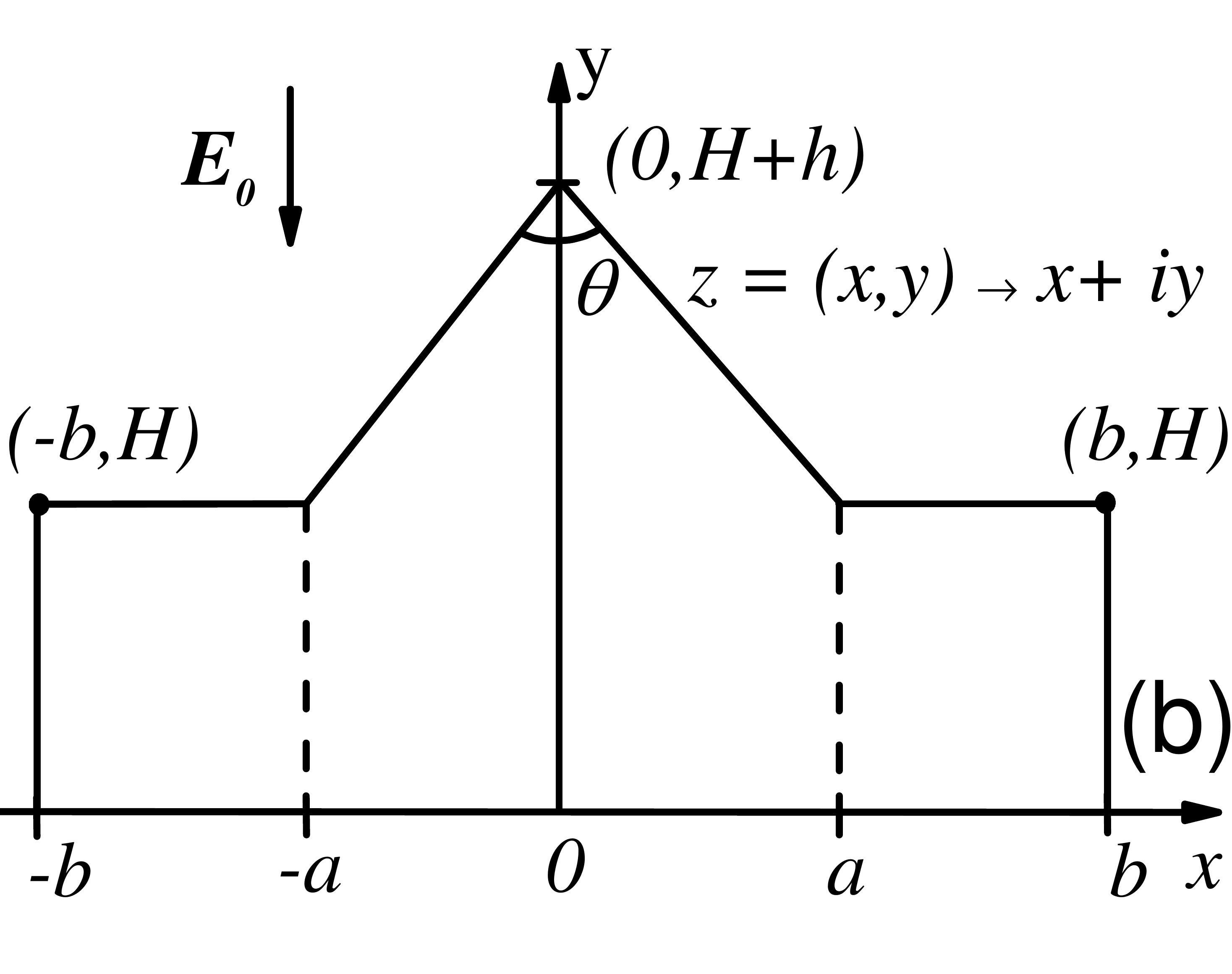}
    \hspace{+0.6cm}
    \includegraphics[width=0.30\linewidth]{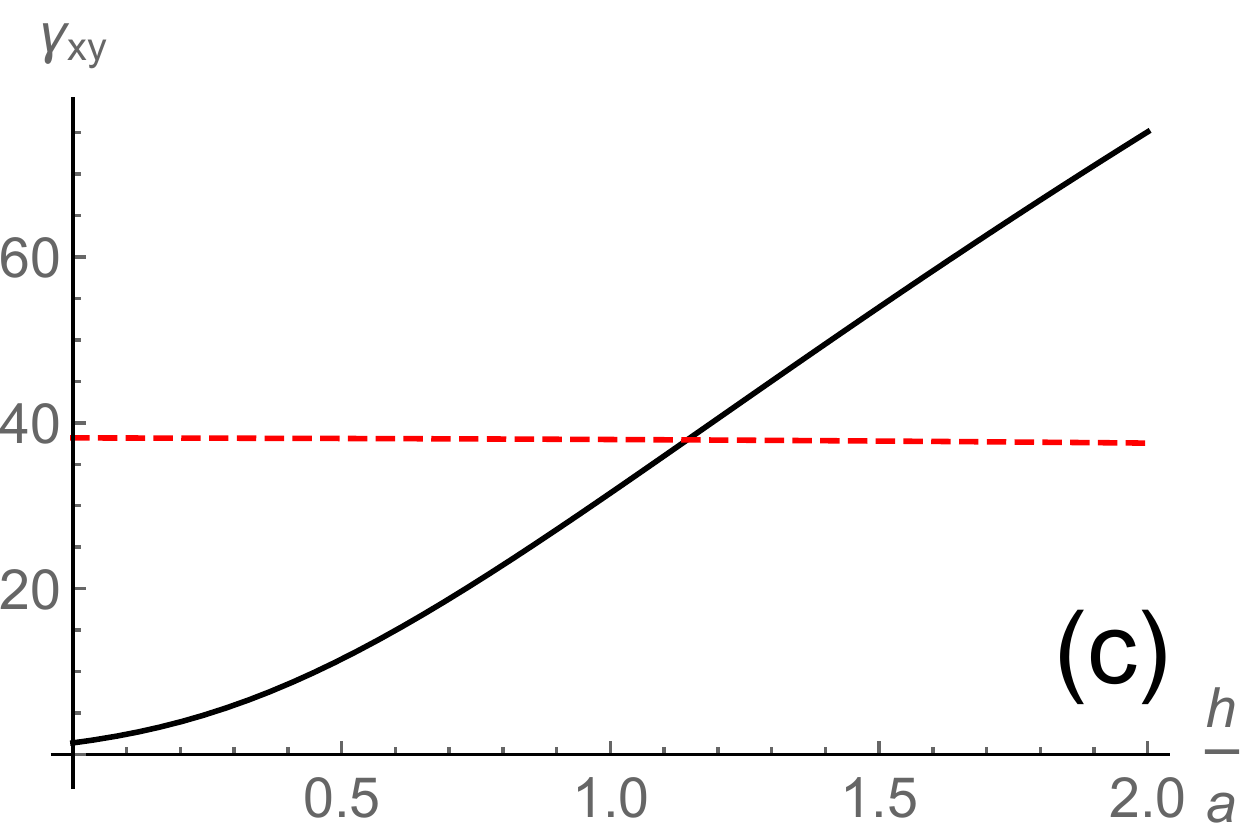}
    \hspace{-0.1cm}
	\caption{(Color online) Conformal mapping between the (a) $w$ and (b) $z$ planes for the two-dimensional problem of a field emitter consisting of a triangular protrusion of height $h$ and half-width $a$, placed on the top of a rectangular protrusion of height $H$ and half-width $b$, on a conducting line. (c) FEF near the rectangular and the triangular corners as a function of the ratio $h/a$ for $H/b=1$, $b/a=10$ and $d/a=0.0001$. The (red) dashed line corresponds to the FEF near the rectangular corner and the (black) full line to the triangular one.}
	\label{twostage}
\end{figure*}

In this section, we consider the two-dimensional problem of a triangular protrusion placed on the top of a rectangular one on a conducting line, when this system is under an external electrostatic field. To do this, we use a conformal transformation between $w$ and $z$-planes, in such a way that we map the horizontal axis in the $w$-plane, showed in Fig.\ref{twostage}(a), to the polygonal line showed in Fig.\ref{twostage}(b). This Schwarz-Christoffel mapping is such that the following requirements are fulfilled: $w=(0,0) \leftrightarrow z=(0,H+h)$, $w=(\pm 1,0) \leftrightarrow z=(\pm a,H)$, $w=(\pm u_{0},0) \leftrightarrow z=(\pm b,H)$, $w=(\pm v_{0},0) \leftrightarrow z=(\pm b,0)$ and is given by Eq.(\ref{SC}), with the function $f$ being expressed by,
\begin{equation} \label{f2}
f(w)=\frac{w^{1-\alpha}}{(w^{2}-1)^{\frac{1-\alpha}{2}}} \sqrt{\frac{w^{2}-u_{0}^{2}}{w^{2}-v_{0}^{2}}}.
\end{equation}

One should notice that we had chosen three specific points in the $w$-plane, corresponding to three of the vertices of the polygonal line in Fig.\ref{twostage}(b). The other points are not determined, since we still don't know $u_0$ and $v_0$. Furthermore, in the previous problem of a triangular protrusion on a line, we could choose all the three points corresponding to the vertices. The possibility of choosing these three arbitrary points in the Schwarz-Christoffel transformation is a particular example featuring the validity of the Riemann Mapping Theorem \cite{Riemann}. This theorem states not only the existence of a bi-holomorphic function, mapping any simply connected, proper and opened subset to the interior of the unit disc, but also states the uniqueness of this function if one fixes one point and the derivative's argument of the transformation on this point.

Choosing $z_{0}=0$ in Eq.(\ref{SC}), we obtain $B=i(H+h)$ from $w=(0,0) \leftrightarrow z=(0,H+h)$. The other requirements lead to the following equations:

\begin{equation}
\tan\left( \frac{\theta}{2} \right)=\frac{a}{h},
\label{E2}
\end{equation}

\begin{equation}
A=\frac{\sqrt{a^{2}+h^{2}}} {\int_{0}^{1} \frac{w^{1-\alpha}}{(1-w^{2})^{\frac{1-\alpha}{2}}} \sqrt{\frac{u_{0}^{2}-w^{2}}{v_{0}^{2}-w^{2}}} dw},
\label{E3}
\end{equation}

\begin{equation}
b-a=A \int_{1}^{u_{0}} \frac{w^{1-\alpha}}{(w^{2}-1)^{\frac{1-\alpha}{2}}} \sqrt{\frac{u_{0}^{2}-w^{2}}{v_{0}^{2}-w^{2}}} dw,
 \label{E4}
 \end{equation}

\begin{equation}
H=A \int_{u_{0}}^{v_{0}} \frac{w^{1-\alpha}}{(w^{2}-1)^{\frac{1-\alpha}{2}}} \sqrt{\frac{w^{2}-u_{0}^{2}}{v_{0}^{2}-w^{2}}} dw.
 \label{E5}
\end{equation}
Equation (\ref{E2}) is obvious from the geometry of Fig.\ref{twostage}(b), while Eqs. (\ref{E3}), (\ref{E4}) and (\ref{E5}) determine the constants $A$, $u_0$ and $v_0$. Near the corner $w=0 \Rightarrow z=i(H+h)$, we can adopt the approximation $f(w) \approx exp \left[\frac{-i \pi(1-\alpha)}{2} \right] \frac{u_0}{v_0} w^{1-\alpha}$ in Eq. (\ref{SC}) to obtain $|z-i(H+h)| \approx \frac{A u_{0}}{v_{0}} \frac{|w|^{2-\alpha}}{2-\alpha}$. Thus, Eq. (\ref{Field1}) under the same approximation leads to the FEF near the top of the structure $(w \rightarrow 0)$:
\begin{equation} \label{Beta2}
\gamma_{xy}=\frac{|\mathbf{E}(x,y)|}{E_0} \approx \frac{v_0}{u_0} \left[ \frac{A u_{0}}{(2-\alpha) v_{0} |z-i(H+h)|}\right]^{\frac{1-\alpha}{2-\alpha}}.
\end{equation}

At last we evaluate the FEF near the rectangular corner ($w \rightarrow u_0 \Rightarrow z \rightarrow b+iH$). For that we notice that $w \rightarrow u_{0} \Rightarrow f(w) \approx \frac{-i \sqrt{2 u_{0}}u_{0}^{1-\alpha}}{(u_{0}^{2}-1)^{\frac{1-\alpha}{2}} \sqrt{v_{0}^{2}-u_{0}^{2}}} \sqrt{w-u_0}$. On the other way, Eq. (\ref{SC}) implies that $z(w)=z(u_{0})+A \int_{u_0}^{w} f(w) dw$. Thus, the correspondence between $w=(u_{0},0)$ and $z=(b,H)$, together with our approximation for $f(w)$ when $w \rightarrow u_0$, leads to $|z-(b+iH)| \approx \frac{2^{3/2} \sqrt{u_0} u_{0}^{1-\alpha} A |w-u_{0}|^{3/2}}{3 (u_{0}^{2}-1)^{\frac{1-\alpha}{2}} \sqrt{v_0^{2}-u_{0}^{2}}}$. Finally, these results and Eq. (\ref{Field1}) can be used to determine the FEF near the rectangular corner $(w \rightarrow u_0)$:
\begin{equation}  \label{Betac}
\gamma_{c}=\frac{|\mathbf{E}(x,y)|}{E_0} \approx\left[\frac{A (u_{0}^{2}-1)^{1-\alpha}(v_{0}^{2}-u_{0}^{2})}{3u_{0} u_{0}^{2(1-\alpha)} |z-(b+iH)|} \right]^{1/3}.
\end{equation}

To complete the evaluation of the FEFs on Eq. (\ref{Beta2}) and Eq. (\ref{Betac}), we must solve Eqs. (\ref{E3}), (\ref{E4}) and (\ref{E5}) to find the reminiscent constants  $A$, $u_0$ and $v_0$. This shall be done during the rest of this section. On the following subsection we will present an analytical solution to these equations and to the FEF near the top of the triangular and the rectangular corners under certain restrictions. In the last subsection we provide the general numerical results.

\begin{figure*}[ht]
	\centering    
    \hspace{-1.0cm}
    \includegraphics[width=0.30\linewidth]{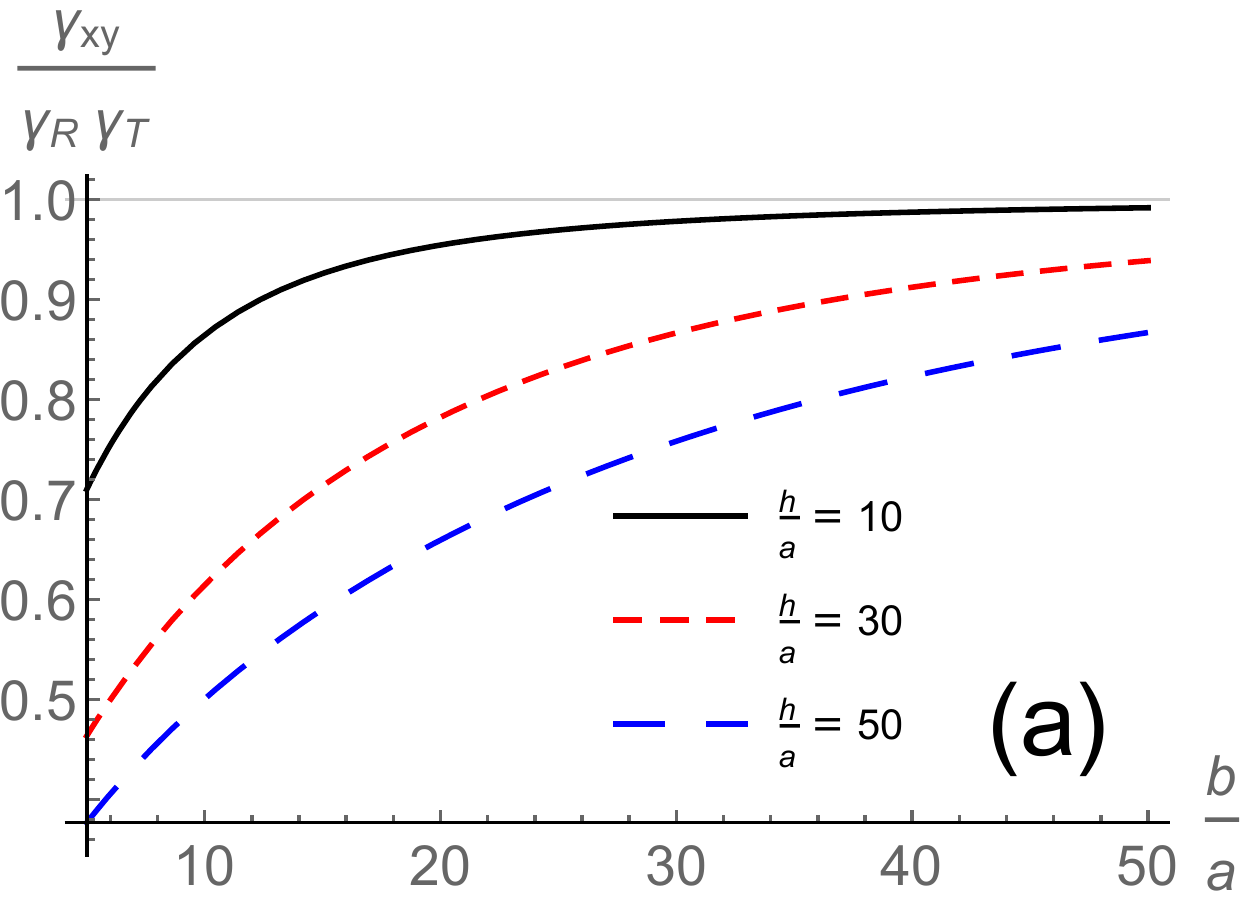}
    \hspace{+0.3cm}
    \includegraphics[width=0.30\linewidth]{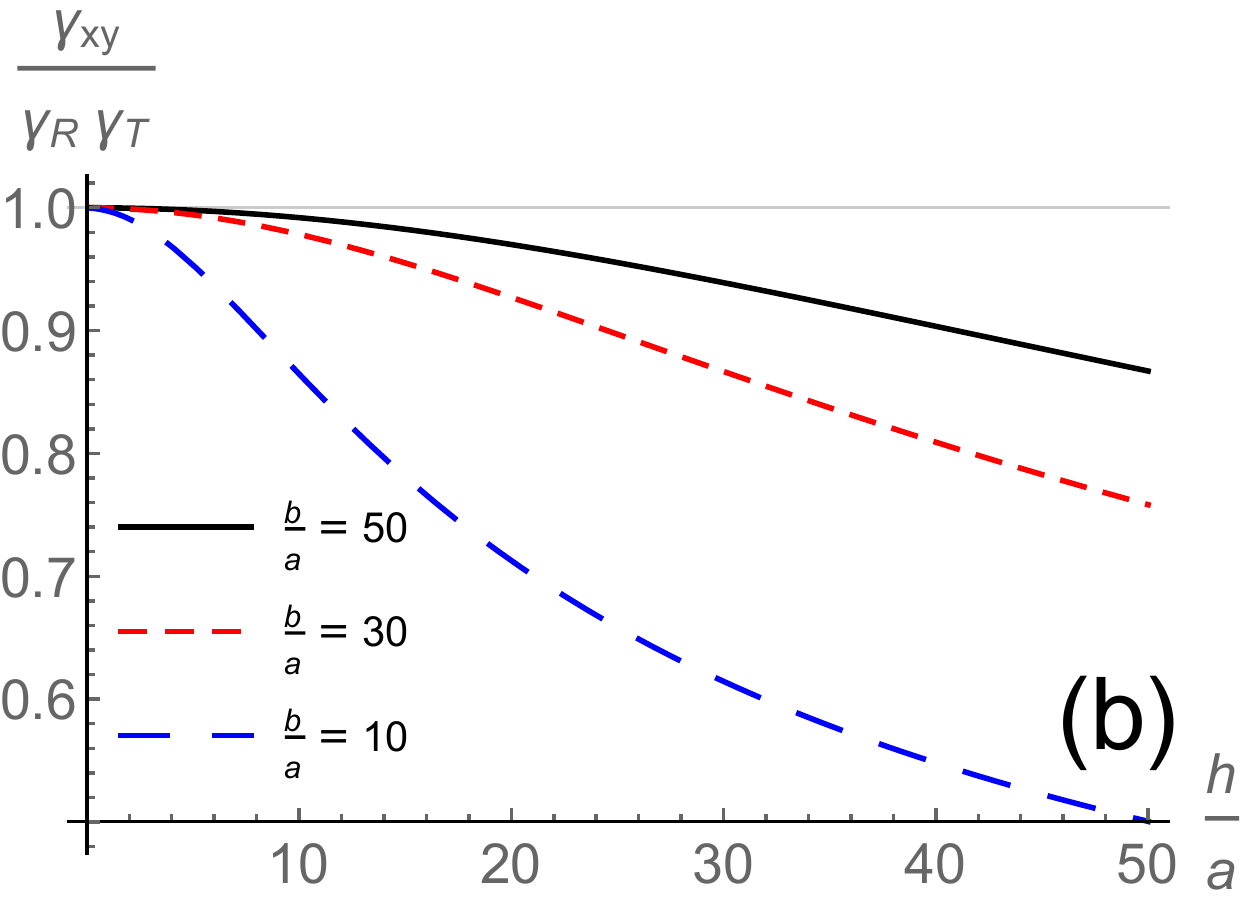}
    \hspace{+0.6cm}
    \includegraphics[width=0.30\linewidth]{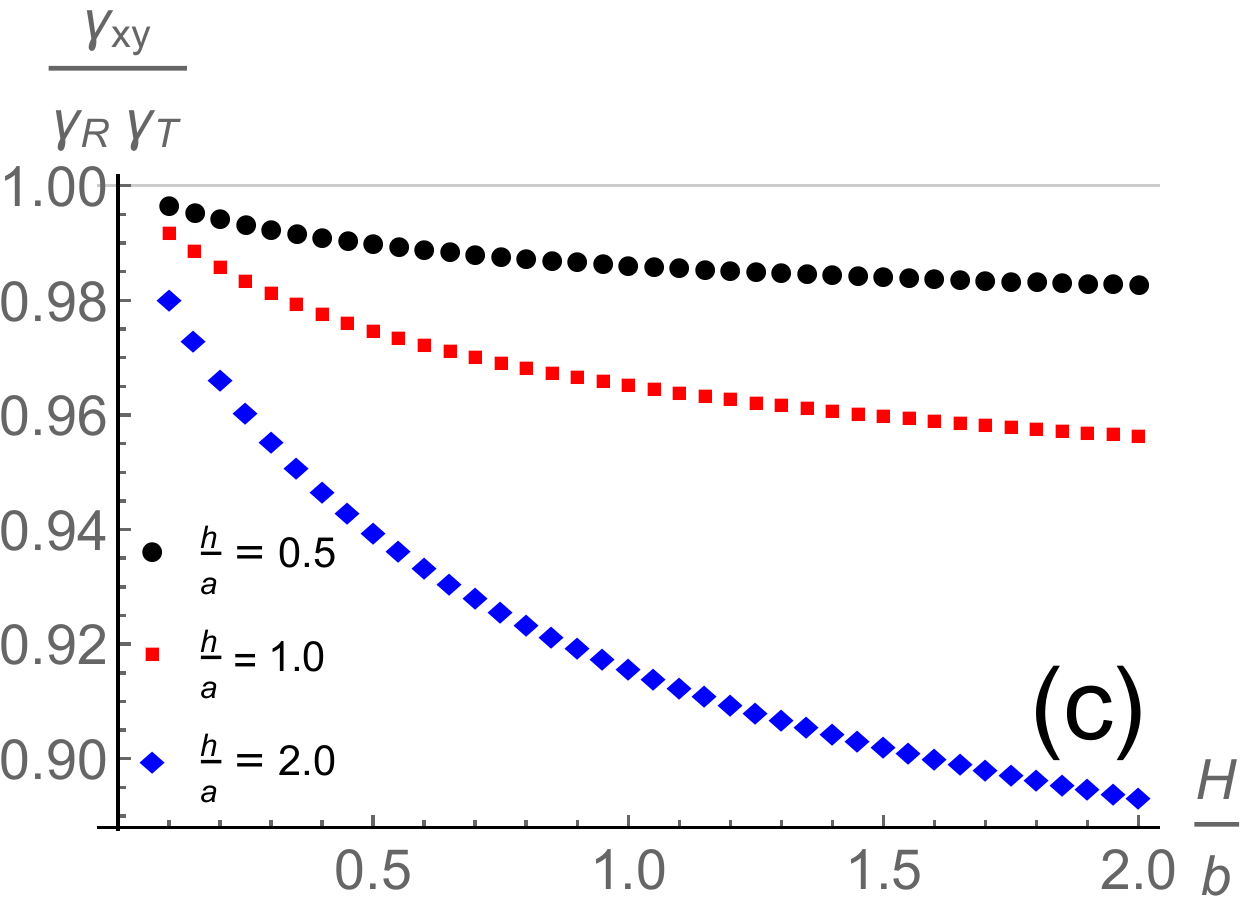}
    \hspace{-0.1cm}
	\caption{(Color online) Ratio between the FEF of the two-stage protrusion and product of the FEFs of each of the stages. In (a) this ratio is showed as a function of $b/a$ for $H/b=50$, $d/a=0.0001$ and different values of $h/a$; in (b) as a function of $h/a$ for $H/b=50$, $d/a=0.0001$ and different values of $h/a$; in (c) as a function of $H/b$ for $b/a=2$, $d/a=0.0001$ and different values of $h/a$.}
 	\label{Schottky}
\end{figure*}

\subsection{Analytical results and analytical proof of Schottky's Conjecture}

Equations (\ref{E3}-\ref{E5}) are too cumbersome for an analytical treatment but this becomes feasible under the limits $\alpha<<1<<u_0<<v_0 \Rightarrow H>>b>>h>>a$. This is an interesting case in which the dimensions of the lower protrusion are much larger than the ones from the other and the height of each protrusion is much larger than its width. Generally, SC \cite{Schottky} is valid in these circumstances, as discussed in Refs. \cite{Ryan1,Ryan2, Jensen,deAssis1} for other geometries.

Under the condition $\alpha<<1<<u_0<<v_0 \Rightarrow H>>b>>h>>a$, Eqs. (\ref{E3}), (\ref{E4}) and (\ref{E5}) become:

\vspace{-0.5cm}
\begin{equation}
A \approx \frac{\sqrt{a^{2}+h^{2}}} { \frac{u_{0}}{v_{0}} \int_{0}^{1} \frac{w^{1-\alpha}}{(1-w^{2})^{\frac{1-\alpha}{2}}}dw} \Rightarrow \frac{A u_0}{v_0} \approx \frac{h \sqrt{\pi}}{\Gamma \left(1-\frac{\alpha}{2} \right) \Gamma \left(\frac{1+\alpha}{2} \right)},
\end{equation}
\vspace{-0.5cm}
\begin{equation}
\frac{b-a}{A} \approx \frac{1}{v_0} \int_{1}^{u_0} \sqrt{u_{0}^{2}-w^{2}} dw \Rightarrow \frac{b}{A} \approx \frac{\pi u_{0}^{2}}{4v_0},
\end{equation}
\vspace{-0.5cm}
\begin{equation}
\frac{H}{A} \approx \int_{u_{0}}^{v_{0}} \frac{w dw}{\sqrt{v_{0}^{2}-w^{2}}}= \sqrt{v_{0}^{2}-u_{0}^{2}} \approx v_{0} \Rightarrow \frac{H}{A} \approx v_0.
\end{equation}

The solution from this non-linear system leads to
\begin{eqnarray}
u_0=\frac{4b \Gamma \left(1-\frac{\alpha}{2} \right) \Gamma \left(\frac{1+\alpha}{2} \right)}{\pi^{3/2} h},  \\
v_0=\frac{2 \sqrt{bH} \Gamma \left(1-\frac{\alpha}{2} \right) \Gamma \left(\frac{1+\alpha}{2} \right)}{\pi h}, \\
A=\frac{\pi h \sqrt{H/b}}{2 \Gamma \left(1-\frac{\alpha}{2} \right) \Gamma \left(\frac{1+\alpha}{2} \right)}.
\end{eqnarray}
Using these results on Eq.(\ref{Beta2}), the FEF near the corner $w=0 \Rightarrow z=i(H+h)$ can be found, so that
\begin{equation} \label{Comp_Beta}
\gamma_{xy} \approx \sqrt{\frac{\pi H}{4b}}  \left[\frac{h \sqrt{\pi}}{(2-\alpha) \Gamma \left(1-\frac{\alpha}{2}\right) \Gamma \left( \frac{1+\alpha}{2} \right) d} \right]^{\frac{1-\alpha}{2-\alpha}},
\end{equation}
where $d \equiv d(x,y) \equiv \sqrt{x^{2}+(y-H-h)^{2}}$ is the distance to the corner. Since $\sqrt{\frac{\pi H}{4b}}$ is the FEF centered on the top of a rectangular protrusion on a line for high aspect ratio \cite{Ryan1,Ryan2}, we see that the result given by Eq.(\ref{Comp_Beta}) is the product of the FEFs of each of the stages. It is important to notice that, here, the FEF from the lower-protrusion is not the local characteristic one, which should be near the corner of the rectangle. Considering the three-dimensional corresponding case as the one with a conical protrusion over a cylindrical one, since the maximum (characteristic) local current density from the lower stage is expected to be near the cylinder's border, the FEF on the center of the top middle of this protrusion is difficult to be extracted directly from experiments. However, SC provides the possibility to derive this quantity indirectly from measurements of the current-voltage characteristics in the vicinity of the apex of both the single-stage conical
protrusion and the two-stage structure (conical over cylindrical protrusion) by using, for example, the Fowler-Nordheim plots \cite{ForbesJordan}.

To support this conclusion, we prove that the largest FEF of the two-stage structure presented here is indeed near the triangular corner. Thus, we also evaluate the FEF near the rectangular corner ($w \rightarrow u_0$), $\gamma_{c}$. Using the analytical solutions obtained for $A$, $u_0$ and $v_0$ in Eq. (\ref{Betac}) under the restriction $\alpha<<1<<u_0<<v_0 \Rightarrow H>>b>>h>>a$, we find:
\begin{equation} \label{FEF_R}
\gamma_{c} \approx \left[\frac{A v_{0}^{2}}{3u_{0} |z-(b+iH)|} \right]^{1/3}  \approx \left[\frac{\pi^{1/2} H^{3/2}}{6 \sqrt{b} d} \right]^{1/3},
\end{equation}
where $d \equiv d(x,y) = \sqrt{(x-b)^{2}+(y-H)^{2}}$ is the distance from the point where we evaluate the FEF to the corner $(b,H)$. We see that Eq. (\ref{Comp_Beta}) yields a larger FEF in the $d \rightarrow 0$ limit when $\alpha<1/2$ by comparing the exponents $\frac{1-\alpha}{2-\alpha}$ and $1/3$ from the power laws in $1/d$ in each expression. Since our analytical evaluation is valid only for  $\alpha<<1$, the largest FEF of the two-stage structure presented here is indeed near the triangular corner in the considered regime.

We stress that the geometries considered in this work feature a linear dependence of the FEF on powers of the ratios height-to-width and distance-to-width with exponents $\frac{1-\alpha}{2-\alpha}$ $\left(0<\alpha<1\right)$, $1/3$ and $1/2$. For very small angles, the  $\alpha$-exponent approaches $1/2$, as obtained in Refs. \cite{Ryan1,Ryan2}. Thus, our systems still exhibit a smaller geometry dependence than the linear one of Ref.\cite{Lau}, for an emitter with a Lorentzian shape.

\subsection{Numerical results}

Finally, with the purpose of exploring the space of parameters characterizing the geometry of the emitter beyond the restrictions proposed in the last subsection, we present the numerical results related to the FEFs of the system with a two-stage protrusion presented in this paper. This is done by using Eqs. (\ref{Beta2}) and (\ref{Betac}), where the constants $A$, $u_0$ and $v_0$ are determined by solving numerically Eqs.(\ref{E3}-\ref{E5}).
Figure \ref{twostage}(c) shows the FEF of the two-stage protrusion near the top triangular corner $(\gamma_{xy})$ and near the rectangular corner $(\gamma_c)$, as a function of the ratio $h/a$ out of the restrictions used in our analytical results $\left(H/b=1, b/a=10, d/a=0.0001 \right)$. Since in this case the upper protrusion is much smaller than the lower one, the FEF of the rectangular corner is almost independent of the aspect ratio of the upper protrusion, which explains the constant value exhibited in Fig.\ref{twostage}(c). On the other way, the FEF near the triangular corner is obviously very sensitive to the the triangle's aspect ratio and may assume larger or smaller values than the FEF near the rectangular corner. This change in the position of the local characteristic FEF, obtained here for ridge emitters, is also present in three-dimensional applications considering low-emittance electron beams \cite{JensenEmittance}.

In Fig.\ref{Schottky}, we show plots of the ratio, $\frac{\gamma_{xy}}{\gamma_{R} \gamma_{T}}$, between the FEF in the vicinity of the apex of the system with a two-stage protrusion $(\gamma_{xy})$ and the product of the FEF near the top of a single-stage triangular protrusion $(\gamma_{T})$ to the FEF centered on the top of a single-stage rectangular one $(\gamma_R)$. According to SC, this ratio should be equal to one. The FEF centered on the top of a rectangular protrusion of height $H$ and half-width $b$ on a line has already been found in \cite{Ryan1,Ryan2} and is given by $\gamma_{R}=1/w_0$, where $w_0$ is the solution of the following equation:
\begin{equation}
\frac{H}{b}=\frac{\int_{w_0}^{1} \sqrt{\frac{w^{2}-w_{0}^{2}}{1-w^{2}}}dw}{\int_{0}^{w_0} \sqrt{\frac{w_{0}^{2}-w^{2}}{1-w^{2}}}dw}.
\end{equation}
In Fig. \ref{Schottky}(a) we consider the results for high aspect ratios of both the protrusions of each of the stages ($H/b=50$ and $h/a \geq 10$). We can see that SC is valid in this case only when the upper protrusion is much smaller than the lower one ($b/a>>1$). Furthermore, the SC is less quickly achieved for higher values of $h/a$ for the values plotted. This may seem to be counterintuitive but, in fact, it just reflects the fact that for higher values of both $h/a$ and $b/a$, the condition $b>>h$ may not be fulfilled. Thus, our numerical results reinforce the analytical ones, as expected.
Fig. \ref{Schottky}(b) is amenable to the same analysis of Fig. \ref{Schottky}(a), but it also has another interesting, although expected, feature: it shows that the SC may be valid for small values of the aspect ratio $h/a$. This means that SC may remain true when the condition $h>>a$, used in our analytical calculations, is violated. This result reinforces the well known idea that SC is in general valid when the dimensions of the lower stage are much larger than the upper stage ones, i.e. the condition of high aspect ratio of the protrusions is superfluous. Results from Fig. \ref{Schottky}(c), on the other way, reveal an unexpected aspect. They show that SC may be valid even when both stages feature dimensions from the same order of magnitude. This is not very common in the literature, except for recent work using point and line charge models \cite{Jensen} and for one previous work considering rectangular and trapezoidal protrusions \cite{Ryan2}. Just to elucidate our claim we consider one specific point of our plot: for $H/b=0.5$, $h/a=0.5$ and $b/a=2$, which means $2H=b=4h=2a$, we have $\frac{\gamma_{xy}}{\gamma_{T} \gamma_{R}}=0.989 \approx1$ for $d=0.0001a$, thus SC is indeed valid in an unexpected region of our space of parameters.

\section{Conclusion}

We have determined exactly, via Schwarz-Christoffel transformation, the local characteristic FEF for a conducting system formed by a triangular protrusion on a line and also for a two-stage system with the morphology of a triangular protrusion centered on the top of a rectangular one on a line. We have provided an analytical proof of SC when the dimensions of the upper-stage structure are much smaller than the lower-stage ones, for large enough aspect ratios and by assuming that the FEF of the lower-stage structure is evaluated at the top middle. Besides that we have also showed, by numerical solution of our exact equations, that although SC is not always true for our system, it may be valid even when both stages have dimensions from the same order. Our numerical results also exhibit change in the position of the local characteristic FEF of our two-stage protrusion, which may be in the triangular or in the rectangular corner depending on the
dimensions of each stage. This feature also appears in applications involving low-emittance devices.

The main results presented in this paper intend to reinforce the validity of SC by means of exact and analytical results obtained via Schwarz-Christoffel conformal mapping. Nevertheless, it is important to notice that there is not a direct technological application from the expressions derived here. This happens because our model considers two-dimensional emitters, which feature a logarithmic dependence of the FEF with the curvature of the edge, instead of a power-law dependence, which is typical of three-dimensional emitters (such as the conical ones). Besides all that, field emission applications involving tips is a great challenge both from experimental and theoretical points of view and we mention two of several reasons. First, great accuracy of the measured electric field is required since the emission current density is very sensitive to that. Second, a small area compared to the whole macroscopic site's size is responsible for the major part of the electron emission. Despite all this difficulty, there are still some attempts to study field emission in tips, using for instance, finite-element \cite{Jenkins} and boundary-element \cite{Hartman} techniques.

Finally, our results suggest that although SC is not an absolute truth, it can be viewed as the explanation for large FEFs in many situations. This aspect reinforces that, when orthodox CFE conditions are satisfied, a careful experimental analysis must be performed, including the measurements of the local current density, in order to state whether SC is the origin of giant FEFs, or not.
\vspace{0.3cm}

\section*{Acknowledgements}
The authors acknowledge the financial support from CNPq and CAPES (Brazilian Agencies).

\end{document}